\documentstyle[11pt]{article}
\renewcommand{\baselinestretch}{1.5}

\def\beq{\begin{equation}}
\def\eeq{\end{equation}}
\def\al{\alpha}
\def\be{\beta}
\def\G{\Gamma}
\def\cO{{\cal O}}
\def\pa{\partial}

\def\cI{{\cal I}}
\def\cR{{\cal R}}
\def\vx{\vec{x}}
\def\br{\bar{r}}
\def\svd{\sqrt{- \nabla^2}}

\begin{document}

\begin{flushright}
IHES/P/93/56, BRX TH-353, IASSNS-HEP-93/67, ADP-93-221/M20
\end{flushright}

\begin{flushright}
November, 1993
\end{flushright}

\vspace{.1in}

\begin{center}
{\Large{\bf Nonsymmetric Gravity has Unacceptable Global Asymptotics}}

\vspace{.2in}
\renewcommand{\baselinestretch}{1}
\small
\normalsize
{\it T. Damour\\
Institut des Hautes Etudes Scientifiques\\
91440 Bures sur Yvette\\
and\\
D.A.R.C., CNRS -- Observatoire de Paris\\
92195 Meudon, France}

\vspace{.2in}

{\it S. Deser \\
School of Natural Sciences\\
Institute for Advanced Study\\
Princeton, NJ 08540\\
and\\
Physics Department\\
Brandeis University\\
Waltham, MA 02254, USA}

\vspace{.2in}

{\it J. McCarthy \\
Department of Physics and Mathematical Physics\\
University of Adelaide\\
Adelaide, SA 5005, Australia}

\end{center}
\renewcommand{\baselinestretch}{1.5}
\small
\normalsize

\vspace{.2in}

{\bf Abstract}

  We analyze the radiative aspects of nonsymmetric
gravity theory to show that, in contrast to General Relativity,
its nonstationary solutions cannot simultaneously exhibit acceptable
asymptotic behavior at both future and past null infinity:
good behavior at future null infinity is only possible through
the use of advanced potentials with concomitant unphysical behavior at
past null infinity.

\vspace{.1in}

\renewcommand{\theequation}{1.\arabic{equation}}
\setcounter{equation}{0}

\newpage

{\bf 1. Introduction}

  Recent examinations \cite{DDM} of the nonsymmetric gravitational
theory (NGT) \cite{Ein,Moff} uncovered a number of
fundamental deficiencies: The gauge invariance present in the linearization
of NGT about flat space cannot be deformed to the full theory, resulting
in vector ghost excitations via curvature coupling.  Also, generic solutions
suffered unacceptable asymptotic behavior at future null infinity ($\cI^+$).

  Our main purpose here is to consider more explicitly the mechanism
underlying asymptotic failure of NGT.
The central criterion for
the NGT field equations to admit physically acceptable solutions
appropriate for the description of bounded radiating systems is,
at a minimum, that the metric have -- in a suitable (Cartesian)
coordinate system -- an expansion in powers of $1/r$ about the Minkowski
flat metric, $\eta_{\mu\nu}$.  The key point
in this requirement is that this expansion be valid not only at spatial
infinity ($r \rightarrow \infty$, $t$ fixed) but, most importantly,
at {\it both} $\cI^+$
($r \rightarrow \infty$, $u = t-r$ fixed) and
$\cI^-$ ($r \rightarrow \infty$, $v = t+r$ fixed).
We shall establish here that acceptable
asymptotic behavior at $\cI^+ \, \, (\cI^-)$ is obtainable,
in particular solutions, only at the cost of diverging physical
quantities at $\cI^- \, \, (\cI^+)$.  We note that the
asymptotic behavior at $\cI^+$ of such particular solutions
\cite{CMT} has recently been claimed to demonstrate the
consistency of the model.  In particular, then, our result
displays the physical flaws in those analyses.

  In Section 2 we present the complete argument, after carefully
defining the problem at hand.  The main result follows from a simple lemma
which implies that requiring good asymptotics at $\cI^+$ forces the use of
advanced potentials in constructing the solution.  In an appendix, a
simplified but faithful analogue vector model is discussed, which
exhibits our main points regarding NGT with a minimum of technicality.

\bigskip
\bigskip

\renewcommand{\theequation}{2.\arabic{equation}}
\setcounter{equation}{0}

{\bf 2. Asymptotic Analysis of the Field Equations}

  For the present purpose it suffices to analyze the NGT field equations in
an expansion in powers of the antisymmetric component
$B_{\mu\nu} = - B_{\nu\mu}$ of the total metric variable
$g_{\mu\nu} \equiv G_{\mu\nu} + B_{\mu\nu}$ \cite{Kelly,DDM}.
Indeed, we will only need the
field equations to first order in $B$ to exhibit the radiative instability
of NGT; the higher order powers can only exacerbate the
problems.  We rewrite the NGT field equations (Eqs. (3) and (4) of \cite{DDM})
as propagation equations for the three independent fields $G_{\mu\nu}$,
$B_{\mu\nu}$ and $\G_\mu$,
\begin{eqnarray}
R_{\mu\nu}(G) &=& \cO (B^2,\G B) \, ,
\label{L1} \\
\Box \G_\nu &=& - 3 \nabla^\mu ( R^{\al\,\,\be}_{\,\,\mu\,\,\nu} B_{\al\be})
+ \cO (B^3, \G B^2) \, ,
\label{L2} \\
\Box B_{\mu\nu} &=& {4 \over 3} (\pa_\mu \G_\nu - \pa_\nu\G_\mu) +
2 R^{\al\,\,\be}_{\,\,\mu\,\,\nu} B_{\al\be} + \cO (B^3, \G B^2) \, ,
\label{L3}
\end{eqnarray}
together with the two subsidiary conditions
\begin{eqnarray}
\nabla^\nu \G_\nu &=& 0 \, ,
\label{L4} \\
\nabla^\nu B_{\mu\nu} &=& \cO (B^3) \, .
\label{L5}
\end{eqnarray}
Here $G_{\mu\nu}$ is the Riemannian metric that acts as the background for
the remaining equations and is used for all covariant operations
({\it e.g.}, to define the wave operator $\Box$); the vector $\G_\mu$,
which originated as a contracted torsion, is a propagating field despite
its role as a Lagrange multiplier ensuring condition (\ref{L5}).  The
constraint (\ref{L4}) represents a gauge choice on $\G_\mu$.  Consistency
between the propagation and subsidiary equations is ensured
by differential identities; once (\ref{L4}), (\ref{L5}) are satisfied
on a Cauchy hypersurface they remain valid by virtue of
(\ref{L1}) --  (\ref{L3}).  These equations may now be solved
order by order in an expansion $g_{\mu\nu} = G^{(0)}_{\mu\nu} +
(G^{(1)}_{\mu\nu} + B^{(1)}_{\mu\nu}) + \, .\, .\, . \quad$ about
a given Einstein space metric $G^{(0)}_{\mu\nu}$.
A point to keep in mind is that, besides the effective source terms
on the RHS of eqs (\ref{L2}) and (\ref{L3}), there will necessarily
appear additional localized source terms representing the
matter couplings of the NGT sector.  The arguments given below
apply to the total effective + matter source terms
for the NGT variables.  We now turn to the discussion of radiative
solutions of these equations.

  In broad terms, a radiative (asymptotically flat) solution in
any classical field theory is one in which all the fields (including
the gravitational one) exhibit a $1/r$ decay at {\it both} future
and past null infinity, {\it i.e.} in which the deviations between the
fields and their (constant) asymptotic values behave, at leading order,
as $f_{out}(u,\theta,\phi)/r$ on $\cI^+$ and
$f_{in}(v,\theta,\phi)/r$ on $\cI^-$.  Roughly speaking\footnote{In the
case of gauge fields with associated conserved charges, one must subtract
their associated static potentials when defining the wave amplitudes
$f_{in}$ and $f_{out}$ then ({\it e.g.} remove the Schwarzschild $M/r$
term when defining the gravitational wave amplitude $f_{ij}^{TT}/r$).},
$f_{out}$ ($f_{in}$) measures the amount of outgoing (incoming)
radiation present in the solution.  Among such general
radiative solutions, classical field theory further selects the
``retarded'' ones which contain no incoming radiation on $\cI^-$
($f_{in}(v,\theta,\phi) = 0$) but only outgoing radiation on $\cI^+$,
associated to the time dependence of some source.  A convenient way
of ensuring that one has selected the physically correct solution is to
consider sources that are stationary for all times $t \leq t_0$ and
turn on after $t = t_0$.  Then all fields (notably the gravitational one)
must be stationary up to the time $t_0$.  We refer to such a solution
as a ``causal'' one.  [When iteratively solving the
field equations by means of propagators and covariant
``spin-projection'' operators, the causal nature of the
solutions is enforced by consistently
using {\it retarded} propagators.  This point is further discussed in
the Appendix, in the context of a simplified model of NGT.  We
show there the equivalence of the spin-projection technique with the
more direct approach used in the text for exhibiting the asymptotic
failure of NGT.]

For General Relativity, radiative solutions \cite{CK}, as well as
more specific causal ones \cite{BD}, are known to
exist and to contain a wave zone in which the metric approaches
(in suitable coordinate systems) its Minkowski
value as $f(t \pm r,\theta,\phi)/r$ at $\cI^\mp$.  Thus in GR, not only
can one exhibit solutions with good radiative behavior at either
$\cI^+$ or $\cI^-$, but this behavior does not entail any
radiation (let alone rising behavior in $r$) at the other null infinity.
In contrast, we will see that NGT can at best be ``tuned'' to obtain good
``radiative" behavior at one end only at the cost of totally unphysical
properties of the geometry at the other.  Already in \cite{DDM} it was
shown for the NGT field equations (\ref{L1} - \ref{L3})
that good fall-off behavior at $\cI^+$
was in fact inconsistent with the retarded solution for $\Gamma_\mu$
in (\ref{L2}), the root of the problem being the slow decay at $\cI^+$
of the effective source term
$\partial_\mu \Gamma_\nu - \partial_\nu \Gamma_\mu$ on the
RHS of Eq.(\ref{L3}).  We now show that if one demands that the NGT
variables fall off at $\cI^+$ faster than their normal rate (in an
attempt to ensure the desired outgoing wave behavior of $G_{\mu\nu}$),
then the retarded solution for $\Gamma_\mu$ must be replaced by the
{\it advanced} one.  More precisely, we demand that when
$r \rightarrow \infty$ with $u = t - r$ fixed,
\beq
\G_\mu = \cO(1/r^{1 + \epsilon}) \, , \quad
R^{\al\,\,\be}_{\,\,\mu\,\,\nu} B_{\al\be} = \cO(1/r^{2 + \epsilon}) \, ,
\quad\quad \epsilon >0 \, .
\label{L6}
\eeq
The consequences of $\Gamma_\mu$ propagating by the {\it advanced}
Green's function are quite drastic:
firstly, at all past times $\Gamma_\mu$ then acts as a time-dependent
source in (\ref{L1}) and (\ref{L3}),
making it impossible to have causal solutions
where the geometry is stationary before the time when the matter
sources turn on; secondly and more globally,
time-reversing the above argument immediately provides a contradiction with
the required physical behavior on $\cI^-$.
The underlying physics is illustrated by a simpler model
deforming Maxwell theory to provide a simplified but faithful version
of NGT; it is given in the Appendix.

  Our demonstration of this result uses the following
simple lemma, well-known in its time-reversed formulation
(see, {\it e.g.}, \cite{Fo}),
which we prove here for the sake of convenient reference:

\bigskip
\noindent

Consider, in an asymptotically flat spacetime, a time-dependent source,
$\rho(x)$, falling off as $\cO(1/r^{2 + \epsilon})$
($\epsilon > 0$) near $\cI^+$.
If $\phi(x)$ is a solution of the inhomogeneous wave equation\footnote{For
simplicity of the following argument we consider a scalar wave equation,
but it of course applies to tensor fields as well.}
\beq
\Box \phi(x) = \rho(x) \, ,
\label{Li1}
\eeq
that falls off as $\cO(1/r^{1 + \epsilon})$ at ${\cal I}^+$,
then it is necessarily the {\it advanced} solution,
\beq
\phi(x) = \int d^4x' G_{adv}(x,x') \rho(x') \, .
\label{Li2}
\eeq

\bigskip
\noindent
The proof of this lemma follows from a straightforward application of
Green's theorem (see, {\it e.g.} \cite{Frie}).
Integrating Green's identity over the spacetime region
$\cR = \cR(x,\br) = \{ x' : t' - t \leq \br \}$, for given $t = x^0$ and
$\br > 0$, we have
\beq
\phi(x) - \int_{\cR}dV'\, G_{adv}(x,x') \rho(x')
= \int_{\partial \cR}d\Sigma'^\mu \, [ \partial'_{\mu} G_{adv}(x,x')\,
\phi(x') - G_{adv}(x,x')\, \partial'_\mu \phi(x')] \, .
\label{Li3}
\eeq
where $dV'$ is the (curved) spacetime volume element, $\partial \cR$ the
boundary of $\cR$, {\it i.e.} the spacelike hypersurface $t' = t + \br$,
and $d\Sigma'^\mu = d\Sigma' \, n'^\mu$ the surface element of $\partial \cR$.
For large enough $\br$ we may replace $G_{adv}$
on $\partial \cR$ by its flat space limit $G_{adv}^{(0)}(x,x')$, where
\beq
G_{adv}^{(0)}(x,x') = -{1\over {2\pi}} \theta(t'-t) \delta((x-x')^2) =
- {1\over {4\pi}} {\delta(t - t' + |\vx - \vx'|) \over |\vx - \vx'|} \, .
\label{Li4}
\eeq
The RHS of (\ref{Li3}) then only has a contribution from the intersection
of the support of $G_{adv}^{(0)}$ with the hypersurface
$\partial \cR$, {\it i.e.}, on the 2-sphere $\br = t' - t = |\vx - \vx'|$.
In fact $(R' \equiv |\vx - \vx'|)$,
\begin{eqnarray}
\phi(x) &-& \int_{\cR}dV'\, G_{adv}(x,x') \rho(x') =
 - \int_{t'=t+\br}{d\Sigma'}\, [\partial'_t G_{adv}(x,x')\, \phi(x') -
G_{adv}(x,x')\, \partial'_t \phi(x')] \nonumber \\
&=& {1\over {4\pi}} \int_{t' = t + \br}d\Omega' dR' \, R'\, [- \partial_{R'}
\delta(t - t' + R')\, \phi(x') - \delta(t - t' + R')\, \partial'_t \phi(x')]
\nonumber \\
&=& {1\over {4\pi}} \int_{R'=\br}d\Omega'\, [\partial_{R'}(R'\phi(x'))
- \partial'_t(R'\phi(x'))]|_{t' = t + \br} \, .
\label{Li5}
\end{eqnarray}
The boundary conditions $\phi = \cO(1/r^{1 + \epsilon})$,
$\rho = \cO(1/r^{2 + \epsilon})$ at $\cI^+$
allow one to take the limit $\br \rightarrow \infty$.  In this limit the
RHS of Eq.(\ref{Li5}) vanishes, and the volume integral on the
LHS converges to the RHS of Eq.(\ref{Li2}), thereby
proving our lemma.

  The demonstration now proceeds by applying our lemma to (\ref{L2}) which
is an inhomogeneous wave equation for $\G_\mu$, taking into account
the boundary conditions (\ref{L6}), the second of which implies the
needed fall-off of the effective source term
$\nabla^\mu(R^{\al\,\,\be}_{\,\,\mu\,\,\nu} B_{\al\be})
= \cO(1/r^{2 + \epsilon})$.
Since $\G_\mu$ is a solution of the inhomogeneous wave equation (\ref{L2}),
decaying faster than $r^{-1}$ at $\cI^+$, the lemma proven above
implies that it is the {\it advanced potential} solution, {\it i.e.}, the
convolution of the total (effective $+$ matter) source with the -- unique --
advanced propagator.  In the time-dependent situation of
interest, this implies that $\G_\mu$ -- and hence the curl,
$\pa_\mu \G_\nu - \pa_\nu \G_\mu$ -- is non-zero at all past times.
But the curl is just the effective source for $B$ in (\ref{L3}),
which must therefore be nonvanishing at all times in the past,
and from (\ref{L1}) we see the same must be true for $G_{\mu\nu}$.
Clearly this is inconsistent with the proposition that the
solution with acceptable asymptotic behavior on ${\cal I}^+$ describes
in a causal manner gravitational radiation associated to a matter source
turned on at some time $t_0$.

  One might attempt to salvage -- if not a causal solution
properly so-called -- at least a geometry whose
good asymptotics on $\cI^+$ can coexist with
good fall-off behavior on $\cI^-$.  However the time-reverse of the
simple argument in \cite{DDM} shows that this is simply not possible.
The assumption of good fall-off for $G$ and $B$, and the discussion above,
imply that the source, $\nabla (R \cdot B) + $ matter contribution,
generates {\it via the
advanced Green's function} (at fastest) a $1/r$ fall-off
for $\Gamma_\mu$ on $\cI^-$.  But that is clearly inconsistent with $1/r$
fall-off for $B_{\mu\nu}$ -- on using (\ref{L3}) we see that at
best $B_{\mu\nu}$ behaves as $r^0$ at $\cI^-$.
This in turn forbids the desired behavior of $G_{\mu\nu}$  when inserted
into (\ref{L1}).
[For, the r.h.s. of (\ref{L1}) contains, {\it e.g.},  $B$ curl $\G$
contributions \cite{DDM} which decrease as $r^{-1}$.]   There are two ways
of verifying the implications we draw about fall-off at $\cI^-$.  The first
is obtained by writing the flat spacetime wave operator in $(v,r,\theta,\phi)$
coordinates: $\Box = 2 ( \partial_r + 1/r)\partial_v +
\partial_r^2 + (2/r) \partial_r + (1/r^2) \Delta_{\theta\phi}$.
This makes it obvious that a source term
$\sim \rho_{n}(v,\theta,\phi)/r^{n}$ implies at fastest
a $\sim \phi_{n-1}(v,\theta,\phi)/r^{n-1}$ fall-off for the solution
(in the special case $n=2$, the solution falls off as $\log(r)/r$ at best).
The second is obtained by noticing that, from Eqs.(\ref{L2}) and (\ref{L3}),
the 2-form $B_{\mu\nu}$ satisfies an equation of the form (with indices
suppressed) $\Delta^2 B = \nabla\nabla(R \cdot B) + \cO(B^3,\Gamma B^2)$,
where $\Delta$ denotes the Hodge-De Rham wave operator.  Like $\Delta$,
the iterated operator $\Delta^2$ is a hyperbolic differential operator which
possesses uniquely defined retarded and advanced Green's functions.
Generalizing the lemma given above, we see that if one demands that
$B$ and $R \cdot B$ fall off sufficiently fast at $\cI^+$, $B(x)$ must
necessarily be given by the convolution of the total source term,
$\nabla\nabla(R \cdot B)$ + matter contribution, with the {\it advanced}
Green's function of $\Delta^2$, say $H_{adv}(x,x')$.  In the flat spacetime
limit the latter Green's function reads
\beq
H^{(0)}_{adv}(x,x') = {1\over {8\pi}} \theta(t' - t - |\vx - \vx'|) \, ,
\label{Ll1}
\eeq
and does not fall off with $|\vx - \vx'|^{-1}$.  The matter contribution
to the source of $\Delta^2$ will generate an advanced $B$ wave behaving
as $r^0$ on $\cI^-$, while the extended effective source can even give a
worse fall-off if $\nabla\nabla(R \cdot B)$ does not decay fast enough at
infinity.   [Let us note for completeness that in the pure vacuum NGT case
our argument must be phrased somewhat differently: good fall-off for
$G_{\mu\nu}$ necessitates rapid fall-off for $\Gamma$ and $B$
(such as (\ref{L6})) at both ends. In turn this rapid fall-off excludes the
presence of both incoming and outgoing NGT radiation (exclusion at one end
only would suffice for the argument to go through).  Then the unique solution
of the hyperbolic system satisfied by $\Gamma$ and $B$ must be the trivial
(pure GR) one, $\Gamma = B = 0$ everywhere.]

  To summarize, we have shown that NGT necessarily displays unphysical
asymptotic behavior in its time-dependent solutions, and in particular that
the asymptotic (but near $\cI^+$ only) solutions proposed
in \cite{CMT} behave catastrophically at past null
infinity.  This loss of correct asymptotic behavior can ultimately be traced
back to the fundamental shortcoming of NGT that it violates the gauge
invariance of the long-range $B_{\mu\nu}$ field.  Indeed, the gauge invariant
(string-generated) massless $B$ theory coupled to Einstein gravity does
not suffer from these defects.  We mention that the other remedy (which
has the further virtue of allowing phenomenologically interesting matter
couplings) is to endow $B$ with a finite range \cite{DDM}.

{\bf Acknowledgements:}

 S.D. has been supported by NSF grants PHY88-04561, 92-45317,  and by
the Ambrose Monell Foundation.  J.M. is supported by the Australian
Research Council, and thanks the IAS for hospitality at the completion
of this work.

\bigskip
\bigskip

\renewcommand{\theequation}{A.\arabic{equation}}
\setcounter{equation}{0}

{\bf Appendix -- A Simple Vectorial Model of NGT}

  We consider here the simplest analogue of the $B$-field system, which
describes a vector field $A_\mu$ coupled to an external nonconserved
localized source $J_\mu$.  The Lagrangian is
${\cal L} = (-1/4) F_{\mu\nu}^2 - \partial^\mu S A_\mu + J^\mu A_\mu$,
while the field equations read
\beq
\partial^\alpha F_{\mu\alpha} =  - \partial_\mu S + J_{\mu} \, ,
\quad \, \quad \partial^\mu A_\mu = 0 \, .
\label{a1}
\eeq
But for the loss of gauge invariance implied by current nonconservation, this
system would simply be a gauge fixing of the Maxwell theory, in which
$S$ enters as a Lagrange multiplier enforcing the condition
$\partial^\mu A_\mu = 0$.  It is
clear that this quasi-Maxwellian vector model closely mimics NGT:
$S$ has the role played by $\Gamma_\mu$ in NGT, while the nonconserved source
$J_\mu$ can either be thought of as replacing the $RB$ term in
(\ref{L3}), or as one of the local matter sources in the
more general case of non-vacuum NGT.  [Note that there is no natural
reason to require conservation of such sources since there
is no corresponding gauge invariance.]

   Despite its origin as a Lagrange multiplier,
$S$ is a relevant excitation because Eq.(\ref{a1}) implies
\beq
\Box S = \partial_\mu J^\mu \neq 0 \, ,
\label{ai1}
\eeq
and it is no longer consistent to set it to zero as it would be in
ordinary gauge-fixed electrodynamics.  In turn, $S$ excites the --
nonpositive energy -- longitudinal modes.  In fact, from the Lagrangian one
easily computes the corresponding energy momentum tensor
\beq
T_{\mu\nu} = F_{\mu}^{\,\,\al} F_{\nu\al} - {1\over 4} \eta_{\mu\nu}
F^{\al\be} F_{\al\be} + \partial_{\mu} S A_{\nu} + \partial_{\nu} S A_{\mu}
- \eta_{\mu\nu} \partial^{\al} S A_{\al}  \, .
\label{a2}
\eeq
The expression for the energy
\beq
T_{00} = {1\over 2} (F_{0i})^2 + {1\over 4} (F_{ij})^2 +
\partial_{0} S A_{0} + \partial_{i} S A_{i}
\label{a5}
\eeq
is clearly not positive definite, reflecting the presence of
negative-energy modes.  Indeed, the canonical analysis for
the $S$/longitudinal sector ({\it i.e.}, under the
temporally local orthogonal decomposition $A_i = A^T_i + A^L_i$, where
$A^L_i = \partial_i a / \svd$) in the absence of sources reveals a
dipole ghost structure \cite{NaLau}: the
action takes the form
\beq
I_0 = \int{d^4x} \, [{1\over 2} (\dot{a} - (\svd A_0))^2 +
S (\dot{A_0} + (\svd a))] \, ,
\label{ai6}
\eeq
and eliminating the $S$-constraint leaves a manifestly dipole-ghost action
for the remaining mode:
\beq
I_{\rm gh} = \int{d^4x} {1\over 2} \Phi \Box^2 \Phi \, , \quad \quad
\Phi(x) = ({1\over \svd} A_0)(x) \, .
\label{ai7}
\eeq

  This fact alone would of course rule out the model as unphysical; however,
the problems in fact occur at the more fundamental level of asymptotics.
  For a local source $J$ with compact support in space time, we may look
for solutions corresponding to purely outgoing radiation in the wave zone.
It is impossible for all the propagating fields to have the usual $1/r$
decay, {\it i.e.}, the retarded form $f(x) = f(t-r)/r$.
The argument here is of course precisely that in the text:
from (\ref{ai1}) for a localized source we find that the retarded solution
for $S$ falls off as $1/r$ at $\cI^+$.
As a consequence the $A_\mu$ wave equation has a source
which falls off too slowly and $1/r$ fall-off for $A$ itself is inconsistent.
As proven in the text, the only way out would be to use advanced
solutions and thereby sacrifice the interpretation as a causal solution,
and more generally to force bad asymptotics at $\cI^-$.

  As a fallacious argument based on covariant spin-projection operators has
been adduced in the last three references in \cite{CMT}, let us indicate here
how our result on the bad fall-off of $A$ can be recovered by introducing the
two orthogonal projection operators $P^T$ and $P^L$ (sufficient in our
quasi-Maxwellian example).  Formally these are
defined  as projection operators such that $P^T + P^L = 1$, where
$P^L_{\mu\nu}(x,x') = \partial_\mu  \Box^{-1}(x,x') \partial'_\nu$.
One should note immediately that, in contrast to the temporally local
projection operators used earlier (which rely only on assuming
the usual {\it spatial} boundary conditions, {\it i.e.}, sufficiently
rapid fall-off at spatial infinity), projection operators based
on $\Box^{-1}$ require global spacetime boundary conditions.
To generate a consistent operator algebra these boundary conditions must be
taken to be the same for all fields, and of course for the present
problem we must impose causal boundary conditions; namely,
all fields and sources must vanish (or become static) before some
time $t_0$.  Then one can consistently define an algebra of projection
operators based on the retarded Green's function $\Box^{-1}_{ret}$
(in mathematical language, such retarded projection operators form
a convolution algebra).  In that case
the longitudinal projection of (\ref{a1}) gives, unambiguously,
\beq
\partial_\mu S(x) = J^L_\mu = \partial_\mu \int{d^4x'}
\, \Box^{-1}_{ret}(x,x') \, \partial'^\nu J_\nu(x') \, ,
\label{ai8}
\eeq
and hence
\beq
S(x) = \int{d^4x'}\, \Box^{-1}_{ret}(x,x')\, \partial'^\nu J_\nu(x') \, .
\label{ai9}
\eeq
The transverse projection of (\ref{a1}) (note that the
second equation in (\ref{a1}) just gives $A_\mu = A^T_\mu$) implies
$\Box A^T_\mu = - J^T_\mu$, whose unique causal solution is
\beq
A^T_\mu = - \int{d^4x'}\, \Box^{-1}_{ret}(x,x')\, J_\mu(x') +
\partial_\mu \int{d^4x'}\, \Box^{-2}_{ret}(x,x')\,
\partial'^\nu J_\nu(x') \, ,
\label{ai10}
\eeq
where the second term involves the retarded version of the
non-decaying Green's function (\ref{Ll1}).
Not surprisingly, we have simply recovered the results discussed above,
with bad fall-off\footnote{Let us recall that in the second
paper of \cite{DDM} we verified explicitly that -- in the
usual matter-coupled version of NGT, considered at the post-Newtonian
approximation -- the analogue of the last term in (\ref{ai10})
(which also contains a divergence) generates non-decaying behavior
for $B_{\mu\nu}$ with the dipole moment of the particle number density
playing the role of a macroscopic source there.} for $A$.  The fallacy in the
arguments of  \cite{CMT} consisted in simply ignoring the crucial,
localized, source terms in (\ref{a1}) and (\ref{ai10}).

  The problems discussed in this simple model serve to elucidate
the fatal flaws undermining the asymptotic -- near $\cI^+$ only --
solutions of \cite{CMT}.  Good asymptotic behavior of $G_{\mu\nu}$
requires rapid $\Gamma$ fall-off; the presence of sources, however,
makes such behavior impossible near $\cI^+$ except by the use of advanced
potentials in solving for $\Gamma$, thus removing the possibility of
acceptable behavior near $\cI^-$.  No consistent causal solutions
are possible in NGT, unlike in GR.


\begin{thebibliography}{99}
\bibitem{DDM}
T. Damour, S. Deser and J. McCarthy, Phys. Rev. {\bf D45}, R3289 (1992);
{\bf D47}, 1541 (1993).
\bibitem{Ein}
A. Einstein, Sitz. Preuss. Akad. der Wiss., 414 (1925);
A. Einstein and E.G. Straus, Ann. Math. {\bf 47}, 731 (1946);
\bibitem{Moff}
J.W. Moffat, Phys. Rev. {\bf D19}, 3554 (1979); and in
{\it Gravitation 1990: A Banff Summer School}, eds. R.B. Mann and P.
Wesson (World Scientific, 1991).
\bibitem{CMT}
N.J. Cornish, J.W. Moffat and D.C. Tatarski, Physics Letters {\bf A 173},
109 (1993); University of Toronto preprint, UTPT-92-17;
N.J. Cornish and J.W. Moffat,
Phys. Rev. {\bf D47}, 4421 (1993); J.W. Moffat, UTPT-93-11 ;
D.C. Tatarski, contributed talk presented at the {\it Fifth Canadian
Conference on General Relativity and Relativistic Astrophysics},
Waterloo, Canada, May 1993.
\bibitem{Kelly}
P.F. Kelly, Class. Quantum Grav. {\bf 8}, 1217 (1991);
{\bf 9}, 1423(Erratum) (1992).
\bibitem{CK}
D. Christodoulou and S. Klainerman, {\it The Global Nonlinear
Stability of the Minkowski Space},
(Princeton University Press,
Princeton, 1993).
\bibitem{BD}
L. Blanchet and T. Damour, Phil. Trans. R. Soc. London {\bf A320}, 379 (1986) ;
L. Blanchet, Proc R. Soc. London {\bf A409}, 383 (1987).
\bibitem{Fo}
V. Fock, {\it The Theory of Space, Time and Gravitation}, (Pergamon Press,
Oxford, 1966) ; see Section 92.
\bibitem{Frie}
F.G. Friedlander, {\it The Wave Equation On a Curved Space-Time}, Cambridge
University Press, Cambridge, 1975.
\bibitem{NaLau}
B. Lautrup, Kgl. Danske Videnskab. Selskab, Mat.-fys. Medd. {\bf 35},
1 (1967) ;
N. Nakanishi, Prog. Theor. Phys. {\bf 35}, 1111 (1966); {\bf 52}, 1929 (1974).
\end{thebibliography}
\end{document}